\documentclass[proceedings]{JHEP3}
\usepackage{cite}

\newcommand{\lwig}{\mbox{\,\raisebox{.3ex}
    {$<$}$\!\!\!\!\!$\raisebox{-.9ex}{$\sim$}\,}}

\PrHEP{PrHEP JHW2002}                   
\conference{Twenty-sixth Johns Hopkins Workshop}                

\usepackage{epsfig}                   

\title{
Ultrahigh energy cosmic rays: clustering, GUT scale and neutrino masses 
}

\author{
\speaker{Z. Fodor$^{a,b}$}\thanks{Invited talk presented at the 26th Johns
Hopkins Workshop on Particle Physics, August 2003, Heidelberg, Germany}
\\ 
$^a$Institute for Theoretical Physics, E\"otv\"os University, Budapest, Hungary\\
$^b$Department of Physics, University of Wuppertal, Germany}

\abstract{
The origin of highest energy cosmic rays (UHECR) is yet unknown.
In order to understand their propagation we determine the 
probability that an ultrahigh energy 
(above $5\cdot 10^{19}$~eV) proton created at a distance $r$ with
energy $E$  arrives at earth above a threshold $E_c$.
The clustering of ultrahigh energy 
cosmic rays suggests that they might be emitted by compact
sources.  A statistical analysis on the source
density based on the multiplicities is presented. 
The ultrahigh energy cosmic ray spectrum is consistent with the
decay of GUT scale particles.
Alternatively, we consider the possibility that a large fraction of the
ultrahigh energy cosmic rays are decay products of Z bosons
which were produced in the scattering of ultrahigh energy cosmic neutrinos on
cosmological relic neutrinos. Based on this scenario
we determine the required mass of the heaviest relic neutrino. 
The required ultrahigh energy neutrino flux should be detected in the near
future by experiments such as AMANDA, RICE or the Pierre Auger Observatory.
}

\begin{document}



\section{Introduction}
The interaction of protons with the microwave 
background predicts a drop in the cosmic ray flux above
the GZK \cite{GZK66} cutoff  $\approx$5$\cdot$10$^{19}$~eV.  
Observations do not confirm this GZK scenario. Instead, 
about 20 events above $10^{20}$~eV
were observed by a number of experiments.
Since above
the GZK energy the attenuation length of particles is a few tens
of megaparsecs if an ultrahigh energy cosmic ray (UHECR) is
observed on earth it was most probably produced in our vicinity.

Any UHECR scenario needs the knowledge of the propagation
of the particles through the cosmic microwave background (CMB).
Section 2 reports on a study \cite{FK00} about
propagation and determines the probability $P(r,E,E_c)$
that protons created at distance $r$ with
energy $E$  reach earth above a threshold $E_c$. Using this $P$
one can give the observed spectrum by one numerical integration
for any injection spectrum.

It is an interesting phenomenon that the UHECR
events are clustered.
Usually it is assumed that
at these high energies the galactic and extragalactic magnetic
fields do not affect the orbit of the cosmic rays, thus they
should point back to their origin within a few degrees. In
contrast to the low energy cosmic rays one can use UHECRs
for point-source search astronomy.
Recently, a statistical analysis \cite{DTT00} based on the multiplicities
of the clustered events estimated the source density.
In Section 3 we we review our extention \cite{FK00} on 
the above analysis. Our analytical approach 
gives the event clustering
probabilities for any space, intensity and energy distribution of
the sources by using a single additional propagation function $P(r,E;E_c)$. 

In Section 4 we review our work \cite{FK01} 
on the scenario that the UHECRs are coming
from decaying superheavy particles (SP) and on the determination 
of their masses  
$m_X$ by an analysis of the observed UHECR spectrum. 
Interestingly enough $m_X$ is compatible with the GUT scale.

The existence of a background gas of
free relic neutrinos is predicted by cosmology.
Ultrahigh energy neutrinos (UHE$\nu$) scatter on 
relic neutrinos (R$\nu$) producing Z bosons, which can decay hadronically 
(Z-burst) \cite{FMS99}. 
In Section 5 we summarize our comparison of 
the predicted proton spectrum with 
observations and review the determination of the mass of the 
heaviest R$\nu$ via a 
maximum likelihood analysis. 

The details of the presented results and a more complete reference list 
can be found in \cite{FK00,FK01,FKR01}. 

\section{Propagation of UHECR protons}

Using pion production as the dominant effect of energy loss for
protons at energies $>$10$^{19}$~eV, ref. \cite{BW99} calculated 
$P(r,E,E_c)$ for three threshold energies. 
We extended \cite{FK00} the results of \cite{BW99}. 
The inelasticity of Bethe-Heitler
pair production is small
($\approx 10^{-3}$), thus we used a continuous energy loss approximation for
this process. The inelasticity of pion-photoproduction is larger
($\approx 0.2 -0.5$) in the energy range of interest, thus there are only a
few tens of such interactions during the propagation. Due to the Poisson
statistics and the spread of the
inelasticity, we will see a spread in the energy spectrum even if the
injected spectrum is mono-energetic.

\begin{figure}\begin{center}
\epsfig{file=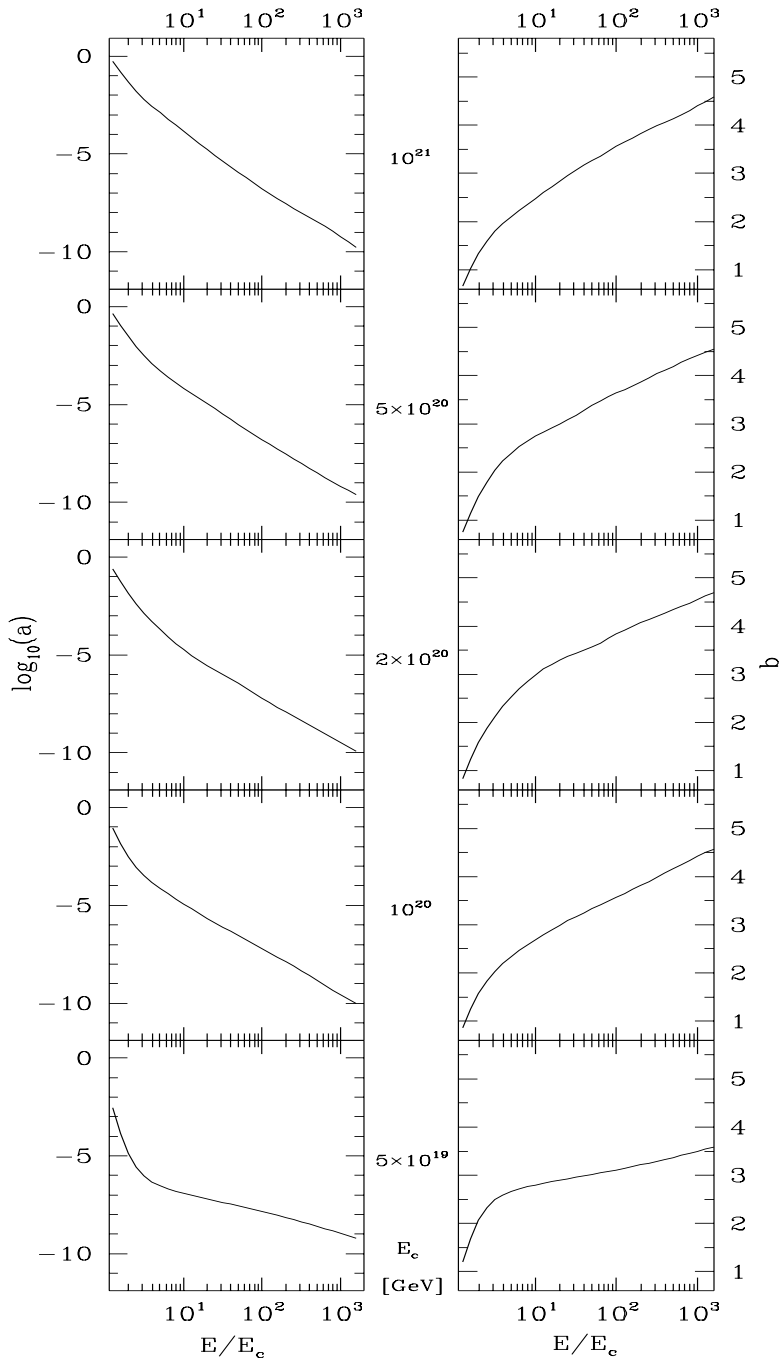,width=12.cm,height=18.0cm,bbllx=210,bblly=180,bburx=460,bbury=700}
\caption{\label{gzk}
{  
The parametrization of $P(r,E,E_c)$ (see Ref. \cite{FK00}).
}}
\end{center}\end{figure}

In our simulation protons are propagated in small steps
($10$~kpc), and after each step the energy losses due to pair
production, pion production and the adiabatic expansion are calculated.
During the simulation we keep track of the current energy of the proton
and its total displacement. 
We used the following type of parametrization
$P(r,E,E_c)=\exp\left[ -a\cdot(r/1\ {\rm Mpc})^b\right]$.
Fig. \ref{gzk} shows  $a(E/E_c)$ and $b(E/E_c)$
for a range of three orders of magnitude and for five different
$E_c$. Just using the functions of $a(E/E_c)$ and
$b(E/E_c)$, thus a parametrization of $P(r,E,E_c)$, one can obtain the
observed energy spectrum for any injection spectrum without additional
Monte-Carlo simulation. 

Since $P(r,E_p;E)$ is of universal usage, we have decided to make the
latest 
numerical data for the probability distribution
$(-)\partial P(r,E_p;E)/\partial E$
available for the public via the World-Wide-Web URL
\begin{center}
{\it http://www.desy.de/\~{}uhecr}  \,.
\end{center}

The propagation function can be similarly determined
for photons \cite{FKR01}, though the necessary CPU 
power is approximately 300 times more
than for protons. Therefore, we used the stochastic method to test a few cases.
Usually the continuous energy loss approximation was used. In this 
approximation, the energy (and number) of the detected photons is a unique
function of
the initial energy and distance, and statistical fluctuations are neglected.
The processes that are taken into account are pair production on the diffuse
extragalactic photon background, double pair production and
inverse Compton scattering of the produced pairs. 
The energy attenuation length of the photons due to these processes
is strongly influenced by the poorely known universal radio
and infrared backgrounds. These uncertainties influences some of our 
results.

A full simulation of the photon propagation function with all the 
statistical fluctuations will be the subject of a later work. 

\section{Density of sources}

The arrival directions of
the UHECRs measured by experiments show some peculiar clustering:
some events are grouped within $ \sim 3^o$, the typical angular
resolution of an experiment. Above $4\cdot 10^{19}$ eV 92 cosmic ray events
were detected, including 7 doublets and 2 triplets.
Above $10^{20}$ eV, one doublet out of 14 events were found \cite{Uchi}.
The chance probability of such a clustering from uniform distribution is
rather small \cite{Uchi,Hea96}. 

The clustered features of the events initiated
an interesting statistical analysis
assuming compact UHECR sources \cite{DTT00}. The authors found
a large number, $\sim 400$ for the number of
sources within the GZK sphere. We generalize their analysis \cite{FK00}.
The most probable value for the source density is really large; 
however, the statistical significance of this result is rather weak. 

The number of UHECRs emitted by a source of luminosity $\lambda$ 
during a period $T$ follows the Poisson distribution.
However, not all
emitted UHECRs will be detected. They might loose their energy during
propagation or can simply go to the wrong direction.
For UHECRs the energy loss
is dominated by the pion production in interaction with
the cosmic microwave background radiation. 
It can be taken into account with the help of the
above mentioned probability
function $P(r,E,E_c)$.

The features of the Poisson distribution enforce us to take
into account the fact that the sky is not
isotropically observed.

The probability of detecting $k$ events from a source at distance
$r$ with energy $E$ can be obtained by simply including the factor
$P(r,E,E_c) A\eta/(4\pi r^2)$ in the Poisson distribution:
\begin{eqnarray}
p_k({\bf x},E,j)
=\frac{\exp\left[  -P(r,E,E_c)\eta j/r^2 \right] }{k!}\times \nonumber\\
\left[ P(r,E,E_c)\eta j/r^2\right] ^k, \label{poiss2}
\end{eqnarray}
where we introduced $j=\lambda T A/(4\pi)$ and $A\eta/(4\pi r^2)$, which is
the
probability that an emitted UHECR points to a detector of
area $A$. The factor $\eta$ represents the visibility of the source,
which was determined by spherical astronomy.
We denote the space, energy and
luminosity  distributions of the sources by $\rho({\bf x})$,
$c(E)$ and $h(j)$, respectively. The probability of detecting $k$
events above the threshold $E_c$ from a single source
randomly positioned within a sphere of radius $R$ is
\begin{eqnarray}\label{P_k}
P_k=\int_{S_R} dV\; \rho({\bf x}) \int_{E_c}^{\infty} 
dE\; c(E) \int_0^{\infty} dj\; h(j) \times \nonumber \\ 
\frac{\exp\left[ -P(r,E,E_c)\eta j/r^2\right] }{k!} \left[
P(r,E,E_c)\eta j/r^2 \right] ^k. \end{eqnarray}

Denote the total number of sources within the sphere of
sufficiently large radius (e.g. several times the GZK radius)
by $N$ and the number of sources that gave $k$ detected events by
$N_k$. Clearly, $N=\sum_0^{\infty}N_i$ and the total number of detected
events is $N_e=\sum_0^{\infty}i N_i$. The probability that for $N$
sources the number of different detected multiplets are $N_k$ is:
\begin{equation}\label{distribution}
P(N,\{N_k\})=N!\prod_{k=0}^{\infty} \frac{1}{N_k!}P_k^{N_k}.
\end{equation}
For a given set of
unclustered and clustered events ($N_1$ and
$N_2,N_3$,...)
inverting the $P(N,\{N_k\})$ distribution function
gives the most probable value for the number

Note, that $P_k$ and then $P(N,\{N_k\})$ are easily determined by
a well behaved four-dimensional numerical integration
for any $c(E)$, $h(j)$ and $\rho (r)$ distribution functions.
In order to illustrate the uncertainties and sensitivities of the
results we used a few different
choices for these distribution functions.

For $c(E)$ we studied three possibilities. The most
straightforward choice is the extrapolation of the `conventional
high energy component' $\propto E^{-2}$. Another possibility is
to use a stronger
fall-off of the spectrum at energies just below the GZK cutoff,
e.g. $\propto E^{-3}$.
The third possibility is to assume
that UHECRs are some decay products of metastable superheavy
particles \cite{Gelmini:2000ds,Crooks:2001jw,Ellis:1990iu,Ellis:1992nb,%
Gondolo:1993rn,BKV97,%
KR98,BS98} or topological defects
\cite{Berezinsky:2000az}.
The superheavy particles decay into quarks and gluons which initiate
multi-hadron cascades through gluon bremstrahlung
\cite{Berezinsky:1998ed,Berezinsky:2001up,Toldra:2002yz,Toldra:2002sn,%
Barbot:2002ep,Ibarra:2002rq}.

In the recent analysis \cite{DTT00}
the authors have shown that for a fixed
set of multiplets the minimal density of sources can be obtained
by assuming a delta-function distribution for $h(j)$. We
studied both this limiting luminosity, $h(j)=\delta(j-j_*)$,
and a more realistic one
with Schechter's luminosity function, which
can be given as:
$h(j)dj=h\cdot (j/j_*)^{-1.25}\exp(-j/j_*)d(j/j_*)$.

The space distribution of sources can be given based on some
particular survey of the distribution of nearby galaxies
or on a correlation length $r_0$ characterizing
the clustering features of sources. For simplicity
the present analysis deals with a homogeneous distribution of
sources.

In order to determine the confidence level (CL) regions for the
source densities we used the frequentist method \cite{PDG}.
We wish to set limits
on S, the source density. Using our Monte-Carlo based
$P(r,E,E_c)$ functions and our analytical technique we
determined $p(N_1,N_2,N_3,...;S;j_*)$, which gives the probability of
observing $N_1$ singlet, $N_2$ doublet, $N_3$
triplet etc. events
if the true value of the density is $S$ and the central value of
luminosity is $j_*$.
For a given set of
$\{N_i,i=1,2,...\}$ the above probability distribution as a
function of $S$ and $j_*$ determines the 68\% and 95\%
confidence level regions in the $S-j_*$ plane.

\begin{figure}\begin{center}
\epsfig{file=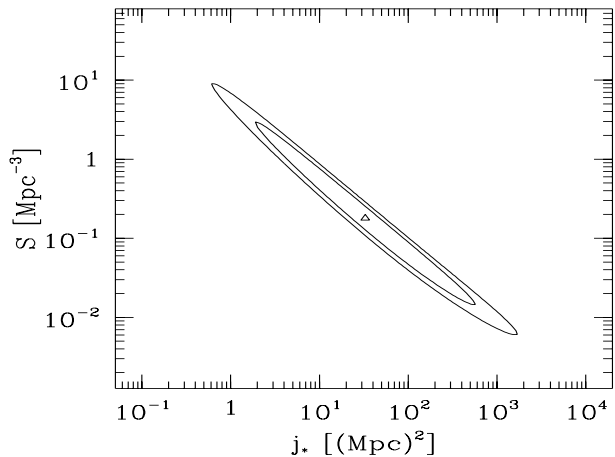,width=7.7cm,height=6.0cm,bbllx=30,bblly=170,bburx=550,bbury=700}
\caption{\label{ell}
{ The $1\sigma$ (68\%) and $2\sigma$ (95\%) confidence
level regions for $j_*$ and the
source density (14 UHECR with one doublet \cite{FK00}). 
}}
\end{center}\end{figure}

Fig. \ref{ell} shows the confidence level regions for one of our
models (with injected energy distribution
$c(E) \propto E^{-3}$; 
and Schechter's luminosity distribution: 
$h(j)dj \propto (j/j_*)^{-1.25} \cdot \exp(-j/j_*)d(j/j_*)$).
The regions are deformed, thin ellipse-like objects.
For this model our final answer for the density is
$180_{-165(174)}^{+2730(8817)}\cdot 10^{-3}$~Mpc$^{-3}$,
where the first errors
indicate the 68\%, the second ones in the parenthesis the 95\%
CLs, respectively.
The choice of \cite{DTT00} --h(j)$\propto\delta$(j)--
and, e.g. $E^{-2}$ energy distribution
gives much smaller value:
$2.77_{-2.53(2.70)}^{+96.1(916)} 10^{-3}$~Mpc$^{-3}$, which is in a quite
good agreement with the result of Ref. \cite{DTT00}.

\section{Decay of GUT scale particles}

An interesting idea discussed by refs.\cite{BKV97,KR98,BS98} is that 
SPs could be the source of UHECRs. 
Note, that any analysis of SP decay
covers a much broader class of possible sources.
Several non-conventional UHECR sources
produce the same UHECR spectra as decaying SPs.
We studied the scenario that the UHECRs are coming
from decaying SPs and we determined the mass of this $X$
particle $m_X$ by a detailed analysis of the observed UHECR 
spectrum \cite{FK01}.

\begin{figure}\begin{center}
\epsfig{file=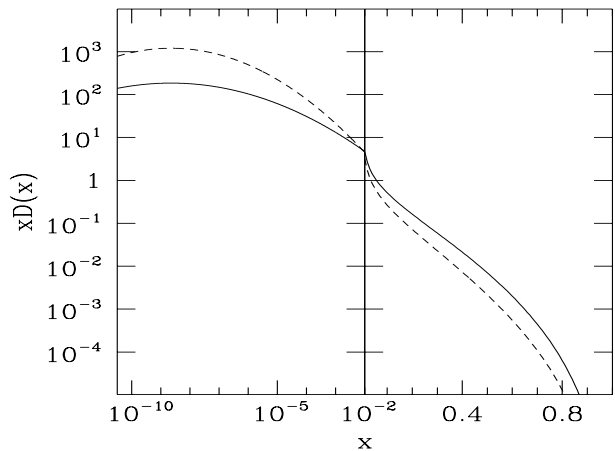,width=7.7cm,height=6.0cm,bbllx=30,bblly=180,bburx=550,bbury=700}
\caption{\label{fragmentation}
{The quark FFs 
at Q=10$^{16}$ GeV for proton/pion in SM (solid/dotted line)  and in MSSM 
(dashed/dashed-dotted line) \cite{FK01}. 
We change from logarithmic scale to linear at $x=0.01$.
}}
\end{center}\end{figure}

The hadronic decay of SPs yields protons. They are characterized 
by the fragmentation function (FF) $D(x,Q^2)$ which gives the number of produced 
protons with momentum fraction $x$ at energy scale $Q$.
For the proton's FF at present accelerator
energies we used ref. \cite{BKK95}. We evolved
the FFs in ordinary and 
in supersymmetric QCD to the energies of the 
SPs. This result can be combined with the
prediction of the MLLA technique, which gives 
the initial spectrum of UHECRs at the energy $m_X$ (cf. Fig.. 
\ref{fragmentation}). Similar results are obtained by
\cite{ST01}. 

Depending on the location of the
source --halo or extragalactic (EG)-- and the 
model --SM or MSSM-- we studied four different scenarios. In the EG case 
protons loose some fraction of their energies, described
by $P(r,E,E_c)$. We compared the predicted and
the observed spectrums by a maximum likelihood analysis.
This analysis gives the mass of the SP and the error on it.

\begin{figure}\begin{center}
\epsfig{file=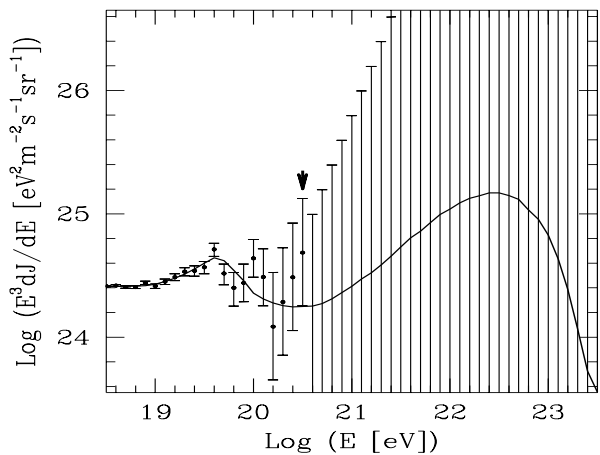,width=7.7cm,height=6.0cm,bbllx=30,bblly=180,bburx=550,bbury=700}
\caption{\label{spect}
{UHECR data with their error bars
and the best fit from a decaying SP \cite{FK01}.
There are no events above $3 \times 10^{20}$~eV 
(shown by an arrow). 
Zero event
does not mean zero flux, but an upper bound for the flux. 
Thus, the experimental 
flux is in the ''hatched'' region with 68\% CL. 
}}
\end{center}\end{figure}

Fig. \ref{spect} shows the measured UHECR spectrum and the best fit, which
is obtained in the EG-MSSM scenario. 

\begin{figure}\begin{center}
\epsfig{file=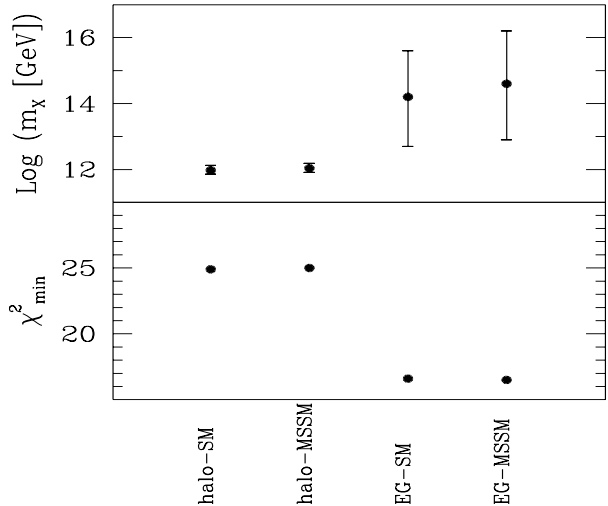,width=8.1cm,height=7.0cm,bbllx=30,bblly=140,bburx=550,bbury=700}
\caption{\label{result}
{The most probable values for the mass of the decaying
ultra heavy dark matter with their error bars and the
total $\chi^2$ values \cite{FK01}. 
Note that 21 bins contain nonzero number of events
and the fit has 3 free parameters.
}}
\end{center}\end{figure}

To determine the most probable value for the mass of the
SP we studied 4 scenarios. Fig. \ref{result} contains
the $\chi^2_{min}$ values and the most
probable masses with their errors for these scenarios.

The UHECR data favors the EG-MSSM scenario. The
goodnesses of the fits for the halo models are far worse.

The SM and MSSM cases do not differ significantly. 
The most important message is that the masses of the best fits 
(EG cases) are compatible within the
error bars with the MSSM gauge coupling unification GUT scale: 
$m_X=10^b$~GeV, where $b=14.6_{-1.7}^{+1.6}$.

\section{Z-burst scenario}

Already in the early eighties 
there were some discussions about the possibility 
that the  ultrahigh energy neutrino spectrum
could have absorption dips at energies around
$E_{\nu_i}^{\rm res} = M_Z^2/(2\,m_{\nu_i}) = 4.2\cdot 10^{21}$ 
(1 eV/$m_{\nu_i}$) eV 
due to resonant annihilation with relic neutrinos of mass $m_{\nu_i}$, 
predicted by the hot Big Bang, 
into Z bosons of mass $M_Z$~\cite{W82,Y97}. 
Recently it was realized that the same annihilation mechanism gives 
a possible solution to the GZK problem~\cite{FMS99}. 
It was argued that the UHECRs above the GZK cutoff are 
from these Z-bursts. The Z-burst hypothesis for the ultrahigh energy cosmic
rays was discussed in many
papers~\cite{W98,Y98,%
GK99,Gelmini:2000ds,%
Weiler:1999ny,Crooks:2001jw,Gelmini:2000bn,Pas:2001nd,Fargion:2000pv,%
FKR01,McKellar:2001hk,%
Kalashev:2001sh,Gelmini:2002xy,Gorbunov:2002nb,Singh:2002de}.

We compared this scenario with observations \cite{FKR01}.

The density distribution of R$\nu$s as hot dark matter follows the total mass
distribution; however, it is less clustered.
To take this into account, the shape of the $n_{\nu_i}(r)$ 
distribution was varied, for distances below 100 Mpc,  
between the standard cosmological homogeneous case 
and that of 
the total mass distribution obtained from peculiar velocity 
measurements. 
These measurements suggest relative overdensities of at most a factor 
$f_\nu =2\div 3$, 
depending on the grid spacing. 
A relative overdensity $f_\nu = 10^2\div 10^4$ in our neighbourhood, as it was 
assumed in earlier 
investigations of the Z-burst 
hypothesis, seems unlikely in view of these data. Note, that our 
quantitative results turned out to be rather insensitive to the 
variations of the overdensities within the considered range.
For scales larger than 100 Mpc the relic neutrino 
density was taken according to the big bang cosmology prediction, 
$n_{\nu_i}=56\cdot (1+z)^3$ cm$^{-3}$. 

We gave the energy distribution of the produced particles 
in our lab system, which
is obtained by Lorentz transforming the CM collider results. 
We included in our analysis the protons, which are 
directly produced in the Z-burst and appear 
as decay products of the neutrons. Photons were also taken into account. They
are produced in hadronic Z decays via fragmentation into neutral pions,
$Z\to \pi^0 + X\to 2\,\gamma + X$. Electrons (and positrons) from hadronic Z
decay are also relevant for the development of electromagnetic cascades.
They stem from decays of secondary charged pions, $Z\to \pi^\pm +X\to e^\pm +
X$.

The UHECR flux from Z-bursts is proportional to the differential fluxes
$F_{\nu_i}$ of ultrahigh energy cosmic neutrinos. Unfortunately,
the value of these fluxes is essentially unknown. In this situation of
insufficient knowledge, we took the following approach concerning the flux of
ultrahigh energy cosmic neutrinos of type $i$, $F_{\nu_i}(E_{\nu_i},r)$. 
It was assumed to have the form
\begin{equation}
F_{\nu_i}(E_{\nu_i},r)=F_{\nu_i}(E_{\nu_i},0)\,(1+z)^\alpha\,,
\end{equation}
where $z$ is the redshift and where $\alpha$ characterizes the cosmological
source evolution. Note, however, that, independently of the production
mechanism, neutrino oscillations result in a uniform mixture for
the different mass eigenstates.

The next ingredient of our analysis is the propagation of the protons and 
photons from cosmological distances. This propagation was
described by the appropriate $P(r,E_p,E)$ probability functions (see Section 2).

Finally, we compared the predicted and observed spectrum and 
extract the mass of the  relic $\nu$ and the necessary UHE$\nu$ flux 
by a maximum
likelihood analysis. 
Qualitatively, our analysis can be understood as follows.
In the Z-burst scenario small relic $\nu$ mass needs large 
$E_\nu^{\rm res}$ to produce a Z. Large $E_\nu^{\rm res}$ results in 
a large Lorentz boost, thus large proton energy. In this way the
detected energy determines the mass of the relic $\nu$. The analysis is completely
analogous to that of the previous section. The observed flux is a sum of two
terms, namely the flux from Z-bursts and a conventional part with power-law
behaviour in the energy. This power-law part might be produced in our galaxy
(halo model) or it might be produced extragalactically (EG model).
 
\begin{figure}\begin{center}
\epsfig{file=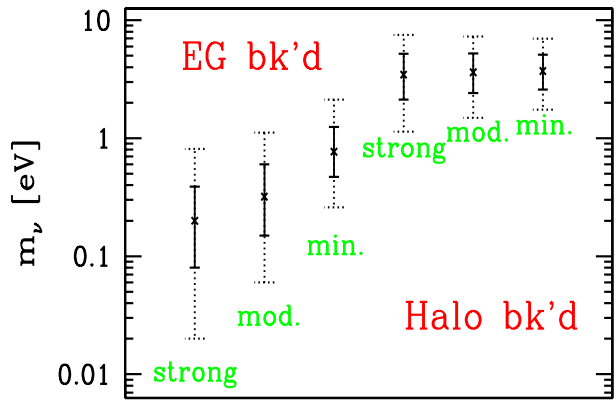,width=7.7cm,height=5.4cm,bbllx=30,bblly=280,bburx=550,bbury=590}
\caption{\label{mass_res}
{
Summary of the masses of the heaviest neutrino required in the Z-burst
scenario, with
their 1\,$\sigma$ (solid) and 2\,$\sigma$ (dotted) error bars, for the case of
an extragalactic and a halo
background of ordinary cosmic rays and
for various assumptions about the diffuse extragalactic photon background in
the radio band
($\alpha =0, h=0.71, \Omega_M= 0.3,\Omega_\Lambda =0.7,z_{\rm max}=2$
see Ref. \cite{FKR01}).
From left: strong $\gamma$ attenuation, moderate and minimal universal radio
background.
}}
\end{center}\end{figure}

The Z-burst determination of the neutrino mass seems reasonably robust.
Fig.~\ref{mass_res} shows the summary of our relic neutrino mass
determination.
For a wide range of cosmological source evolution ($\alpha =-3\div 3$),
Hubble parameters $h=0.61\div 0.9$, $\Omega_M$, $\Omega_\Lambda$, 
$z_{\rm max}=2\div 5$, for variations of the possible relic neutrino
overdensity in our GZK zone 
and for different assumptions about the diffuse extragalactic photon
background,  
the results remain within the above error bars. The main uncertainties
concerning the 
central values originate from the different assumptions about the background
of ordinary
cosmic rays. 

In the case that the ordinary cosmic rays above $10^{18.5}$~eV
are protons and
originate from a region within the GZK zone of about 50~Mpc (``halo''), the
required mass of the
heaviest neutrino seems to lie between
2.1~eV$\le m_\nu \le$6.7~eV
at the 68\,\% C.L. ($\alpha\leq 0$), if we take into account the
variations between the
minimal and moderate universal radio background cases and the 
strong UHE$\gamma$ attenuation case.

The much more plausible assumption that the ordinary cosmic rays above
$10^{18.5}$~eV are
protons of extragalactic origin leads to a required neutrino mass of
0.08~eV$\le m_\nu \le$1.3~eV
at the 68\,\% C.L. ($\alpha\leq 0$). In this case the predicted mass has a
relatively strong dependence on the value of the universal radio background.
Physically it is easy to understand the reason. The small radio background 
leads to a relatively large UHE$\gamma$ fraction in the observations. They
do not loose that much energy. Thus, smaller
incoming UHE$\nu$ energy and larger $m_\nu$ is needed to describe the data. 

We performed a Monte Carlo analysis studying higher statistics. In the near
future, the Pierre Auger Observatory will
provide a ten times higher statistics, which
reduces the error bars in the neutrino mass to about one third of their
present values.

\begin{figure}\begin{center}
\includegraphics[bbllx=20pt,bblly=221pt,bburx=570pt,bbury=608pt,width=8.65cm]
{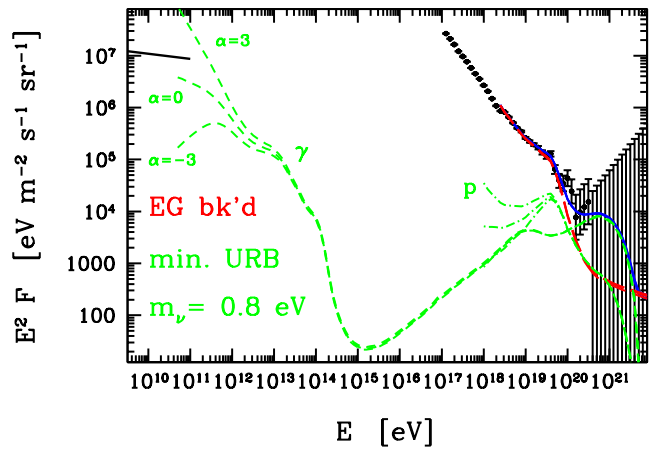}
\caption[...]{\label{fit_egret}
The available UHECR data with their error bars
and the best fit from Z-bursts, 
for various cosmological evolution parameters $\alpha$ and 
an energy attenuation of photons exploiting  
a ``minimal'' URB  
($h=0.71, \Omega_M= 0.3,\Omega_\Lambda =0.7,z_{\rm max}=2$ 
see Ref. \cite{FKR01}).  
Also shown is the diffuse $\gamma$ background in the energy range 
between 30 MeV and 100 GeV as measured by EGRET (solid). 
}
\end{center}
\end{figure}

Let us consider in detail the $\gamma$ ray spectra from Z-bursts, 
notably
in the $\sim 100$~GeV region. As Fig.~\ref{fit_egret} illustrates, 
the EGRET measurements of the diffuse $\gamma$ background in the energy range 
between 30~MeV and 100~GeV~\cite{Sreekumar:1998} gives a strong constraint
on the evolution parameter $\alpha$. 
The high energy spectrum, and thus the neutrino mass,
is independent of $\alpha$, at low energies only
$\alpha\lwig 0$ seems to 
be compatible with the EGRET measurements, quite 
independently of different 
assumptions about the universal radio background (URB).      
These numerical findings are in fairly good agreement with other recent 
simulations~\cite{Kalashev:2001sh}. These photon fluxes do not contain any
contribution from direct photons emitted by the UHE$\nu$ sources. As they are
already close to the EGRET limit, one needs special sources that do not give 
contribution to the EGRET region.

\begin{figure}\begin{center}
\epsfig{file=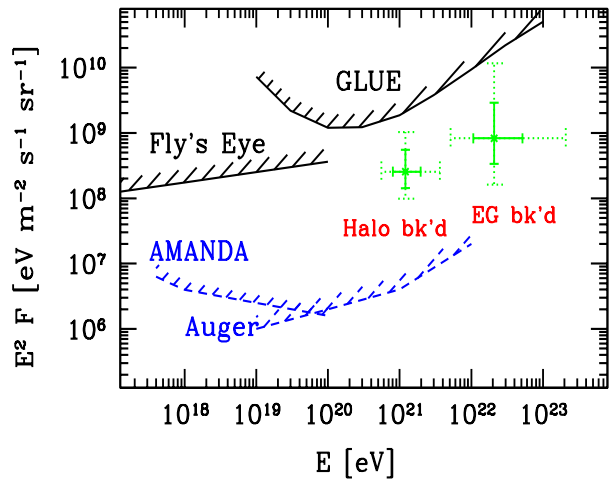,width=7.7cm,height=6.1cm,bbllx=30,bblly=230,bburx=550,bbury=590}
\caption{\label{eflux}
Neutrino fluxes,
$F = \frac{1}{3} \sum_{i=1}^3 ( F_{\nu_i}+F_{\bar\nu_i})$, required
by the Z-burst hypothesis for the case of a halo and an extragalactic
background of
ordinary cosmic rays, respectively ($\alpha =0, h=0.71, \Omega_M=
0.3,\Omega_\Lambda =0.7,z_{\rm max}=2$).
Shown are the necessary fluxes obtained from the fit results 
for the case of a strong UHE$\gamma$ attenuation.
The horizontal errors indicate the 1\,$\sigma$ (solid) and 2\,$\sigma$
(dotted) uncertainties of the
mass determination and the vertical errors include also the uncertainty
of the Hubble expansion rate.
Also shown are upper limits from Fly's Eye on
$F_{\nu_e}+F_{\bar\nu_e}$ and the
Gold\-stone lunar ultrahigh energy neutrino ex\-pe\-ri\-ment
GLUE on $\sum_{\alpha = e,\mu} (
F_{\nu_\alpha}+F_{\bar\nu_\alpha})$,
as well as projected sen\-si\-tivi\-ties of
AMAN\-DA on $F_{\nu_\mu}+F_{\bar\nu_\mu}$
and Auger on $F_{\nu_e}+F_{\bar\nu_e}$. The
sensitiviy of RICE is comparable to the one of Auger.}
\end{center}\end{figure}

It should be stressed that, besides the neutrino mass, the UHE$\nu$ flux at
the resonance
energy is one of the most robust predictions of the Z-burst scenario which can
be verified
or falsified in the near future. The required flux of ultrahigh energy cosmic
neutrinos near the resonant energy should be detected in the near future
by AMANDA, RICE, and the Pierre Auger Observatory, otherwise the Z-burst
scenario will be ruled out (cf. Fig. \ref{eflux}). 
If such tremendous fluxes of ultrahigh energy
neutrinos are indeed found, one has to deal with
the challenge to explain their origin. It is fair to say, that at the moment
no convincing astrophysical
sources are known which meet the requirements for the Z-burst hypothesis, i.e.
which  
have no or a negative cosmological evolution,
accelerate protons at least up to $10^{23}$~eV, are opaque to primary
nucleons and emit secondary photons only in the sub-MeV region.
It is an interesting question whether such challenging conditions can be
realized in BL Lac objects, a
class of active galactic nuclei for which some evidence of zero or negative
cosmological evolution has been
found (see Ref.~\cite{Caccianiga:2001} and references therein) and which were
recently discussed as
possible sources of the highest energy cosmic rays~\cite{Tinyakov:2001nr}.

Z.F thanks the organisers of the 26th Johns Hopkins Workshop
on Current Problems in Particle Physics for the stimulating
atmosphere. 
This work was partially supported by Hung. Sci.
grants No. 
OTKA-T37615\-T34980/\-T29803/\-M37071\-M28413/\-OM-MU-708/\-OMFB-1548.

\end{document}